\documentclass[twocolumn, showpacs, amsmath,amsfonts,prl]{revtex4}
\usepackage{graphicx}
\begin{document}

\title{Thermodynamics of second phase conductive filaments}
\author{V. G. Karpov} \affiliation{Department of Physics and Astronomy, University of Toledo, Toledo, OH 43606, USA}
\author{M. Nardone} \affiliation{Department of Physics and Astronomy, University of Toledo, Toledo, OH 43606, USA}
\author{M. Simon} \affiliation{Department of Physics and Astronomy, University of Toledo, Toledo, OH 43606, USA}
\date{\today}

\begin{abstract}

We present a theory of second phase conductive filaments in phase transformable systems; applications include threshold switches, phase change memory, and shunting in thin film structures. We show that the average filament parameters can be described thermodynamically. In agreement with the published data, the predicted filament current voltage characteristics exhibit negative differential resistance vanishing at high currents where the current density becomes a bulk material property. Our description is extendible to filament transients and allows for efficient numerical simulation.

\end{abstract}
\maketitle

A general observation common to multiple modern technologies is that thin-film structures can drastically decrease their transversal electric resistance by forming conductive filaments (CF) in response to electric bias. CF can be either stable, as in phase change memory and dielectric oxides after hard breakdown, or unstable (disappearing after the bias is removed), as in threshold switches \cite{ovshinsky1968}.  From a practical perspective, CF can cause detrimental shunting and loss of functionality in devices such as thin-film photovoltaics and thin oxides of electronic devices \cite{karpov2004, alam2002}.  On the other hand, in implementations such as threshold switches \cite{adler1978, kau2009}, phase change memory \cite{bedeschi2009}, and resistive memory \cite{chang2009}, CF facilitate information storage and logic operations.

Despite a long history of observations, a theoretical framework for CF does not exist.  Here, we introduce a general thermodynamic theory that describes CF coupled with an external circuit and predicts the CF radius as a function of the electric current and material parameters, as well as the corresponding current-voltage (IV) characteristics.
A finite element numerical simulation is employed to support our analytical results, which are found to be in excellent agreement with the data.

For specificity we consider  the archetypal CF system of chalcogenide glass threshold switches wherein reversible switching takes place between highly resistive (amorphous) and conductive (CF) phases.  They have recently regained interest in connection with 3D stackable phase change memory \cite{kau2009}.  The empirical fact that the CF radius increases with current as $r\propto I^{1/2}$ has been known \cite{adler1978, petersen1976} since the seminal work by Ovshinsky \cite{ovshinsky1968}, yet it still lacks theoretical description.  An early approach based on the principle of least entropy production \cite{ridely1963} did not result in verifiable predictions;  its validity remains questionable \cite{ross2008} and avoidance of it leads to different results, as shown here.

Whether electronic \cite{adler1979, redaelli2008} or crystalline \cite{nardone2009}, or otherwise different from the host material, CF represents a domain of different phase, thus calling upon the analysis of phase transformations.  Our conservative approach avoids the principle of least entropy production starting instead with the kinetic Fokker-Planck equation, the applicability of which to phase transformations, particularly to nucleation phenomena (according to Zeldovich theory), is well established \cite{landau2008}.  Following that approach (see e.g. p. 428 in Ref. \cite{landau2008}) the Fokker-Planck equation in the space of cylinder radii $r$ takes the form,
\begin{equation}\label{eq:FP}
\frac{\partial f}{\partial t}=-\frac{\partial s}{\partial r},\quad s\equiv -B\frac{\partial f}{\partial r}+Af=-Bf_0\frac{\partial }{\partial r}\left(\frac{f}{f_0}\right).
\end{equation}
Here, $f$ is the distribution function so that $f(r)dr$ gives the concentration of filaments in the interval $(r,r+dr)$; $s$ is the flux in radii space (s$^{-1}$ cm$^{-3}$). The `filament radius diffusion coefficient' $B$ can be estimated as $\nu a^2\exp(-W_a/kT)$ where $\nu$ is the characteristic atomic frequency ($\sim 10^{13}$ s$^{-1}$), $a$ is the characteristic interatomic distance, $W_a$ is the kinetic phase transformation barrier, $k$ is Boltzmann's constant, and $T$ is the temperature.  $A$ is connected with $B$ by a relationship which follows from the fact that $s=0$ for the equilibrium distribution $f_0(r)\propto \exp[-F(r)/kT]$, where $F$ is the free energy.

We note that the concept of free energy $F$ that appears with the {\it equilibrium} distribution $f_0$ is not not compromised by the fact that electric current flows through the CF, since that current is fixed by the external circuit and serves only as a temperature source. Hence, $F$ describes the  free energy of the CF in an insulating host {\it parametrically} dependent on the electric current.

The boundary condition $f(r=0)=0$ to Eq. (\ref{eq:FP}) reflects the fact that very thin filaments cannot exist due to limitations such as loss of conductivity or mechanical instability (extraneous to the present model). Another condition, $f(r=\infty )=0$, implies that only finite radii are achievable over finite times $t$.

Using the right-hand-side expression for $s$, multiplying Eq. (\ref{eq:FP}) by $r$, integrating from 0 to $\infty$ by parts, and noting that $\int frdr=\langle r\rangle$, yields $\partial \langle r\rangle /\partial t=\langle\partial F/\partial r\rangle$. We then approximate $\langle F\rangle = F(\langle r\rangle )$ and $\langle\partial F/\partial r\rangle=\partial \langle F\rangle /\partial \langle r\rangle$, thereby neglecting fluctuations in the ensemble of nominally identical filaments. Omitting for brevity the angular brackets, one finally obtains,
\begin{equation}\label{eq:mobility}
\frac{\partial r}{\partial t}=-b\frac{\partial  F}{\partial r}\quad {\rm with}\quad b=\frac{B}{kT}.
\end{equation}
This equation which expresses the average evolution of CF cylinder radius has the standard meaning of a relation between the (growth) velocity and the (thermodynamic) force $-\partial F/\partial r$, with the mobility $b$ and the diffusion coefficient $B$ obeying the Einstein relation.

We limit the present discussion to the steady state case $\partial r/\partial t=0$, which, according to Eq. (\ref{eq:mobility}) takes place when the $F$ is a minimum (obviously different from the condition of least entropy production \cite{ridely1963}). However, in principle, Eq. (\ref{eq:mobility}) is capable of describing various transients.

\begin{figure}[htb]
\includegraphics[width=0.48\textwidth]{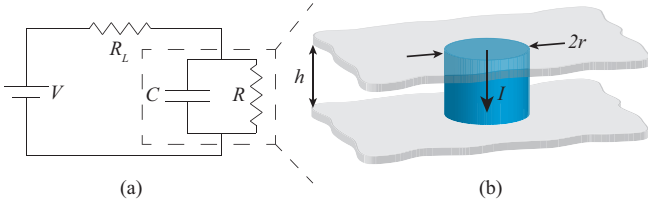}
\caption{(color online) Analytical model components with: (a) circuit schematic showing source voltage $V$, load resistance $R_L$, capacitance $C$, and filament resistance $R$; and (b) flat-plat capacitor of height $h$, and filament of radius $r$ carrying the current $I$.\label{fig:model}}
\end{figure}

To analytically present the free energy we consider a model in Fig. \ref{fig:model} based on a flat plate capacitor of area $A$ and thickness $h$ containing a cylindrical CF of radius $r$. Also, we will assume the characteristic filament dimensions well above the screening length, thereby neglecting possible effects due to electric charge redistribution around the CF.  Including these effects would result in additional terms to the free energy adding mathematical complexity while not changing the approach of this work.

With the above in mind, the major part of the free energy is given by,
\begin{equation}\label{eq:freeen}
F=c\delta T\pi r^2h+\frac{CU^2}{2}+2\pi rh\sigma +\pi r^2h\mu.
\end{equation}
Here, $c$ is the specific heat of the CF, $\delta T$ is the current dependent temperature change, $U$ is the voltage across the capacitor, $C=\varepsilon A/4\pi h$ , $\varepsilon$ is the dielectric permittivity, $\sigma$ is the surface energy, and $\mu$ is the change in chemical potential between the two phases.  The first term in Eq. (\ref{eq:freeen}) represents the thermal contribution, the second is the electrostatic energy, and the last two correspond to the phase transformation.

For the remainder of this analysis we specify the free energy to the case of CF radii $r>h$ because: (1) published experimental results are available; (2) CF dimensions correspond to the above assumed largeness compared to the screening length; and (3) the typical values of the parameters enable one to neglect the second and the third terms in Eq. (\ref{eq:freeen1}), which allows analytical solutions.  In that case, the parameters in Eq. (\ref{eq:freeen}) are,
\begin{equation}\label{eq:deltaT}
U=\frac{VR}{R+R_L}, \quad R=\frac{\rho h}{\pi r^2}, \quad \delta T=\frac{I^2h^2\rho}{8\pi ^2\kappa c r^4},
\end{equation}
where $V$ is the source voltage, $R_L$ is the load resistance, $I$ is the current, $R$ and $\rho$ are the filament resistance and resistivity, respectively, and $\kappa$ is the thermal diffusivity, taken to be the same for the filament and host materials.

Substituting Eqs. (\ref{eq:deltaT}), the free energy becomes
\begin{equation}\label{eq:freeen1}
F= \frac{3Wh}{2r_0}\left\{\frac{\beta x^2}{(1+Hx^2)^2}+\frac{\gamma }{(1+Hx^2)^2}+x+ x^2\right\},
\end{equation}
where $x\equiv r/r_0$ and we have introduced the dimensionless parameters,
\begin{equation}
\beta =\frac{\pi r_0^3V^2}{12W\kappa \rho}, \quad \gamma = \frac{r_0}{h}\frac{CV^2}{3W}, \quad H=\frac{R_L\pi r_0^2}{\rho h}.
\label{eq:gammabeta}\end{equation}
Here the characteristic energy and length,
\begin{equation}\label{eq:Wr}
W=16\pi\sigma ^3/3\mu ^2\quad {\rm and}\quad r_0=2\sigma /\mu .
\end{equation}
would have the physical meaning of nucleation barrier and radius in classical nucleation theory (in which $\mu$ is negative and $|\mu |$ is used instead). Assuming $\sigma$ and $|\mu |$ to be of the same order of magnitude as for crystal nucleation in chalcogenide glasses, one can use the corresponding estimates \cite{nuclparam} $W\sim 2$ eV and $r_0\sim 3$ nm.

\begin{figure}[hbt]
\includegraphics[width=0.48\textwidth]{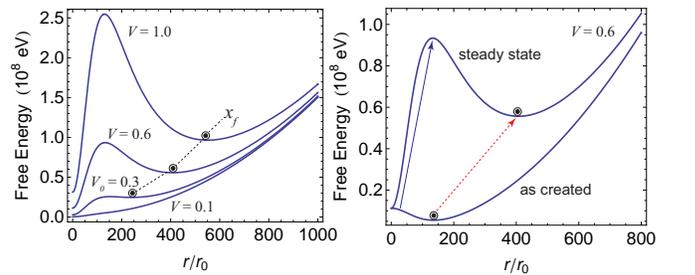}
\caption{(color online) (a) Typical free energy vs. filament radius $x=r/r_0$ at various source voltages $V$ [from Eq. (\ref{eq:freeen1})]. The steady state radius is $x_f$ and CF are unstable at $V<V_0$.  (b) Free energy of a filament as created (thermal contribution neglected) and in the steady state. Arrows show how the energy minimum moves to the right and becomes metastable, separated by a barrier from the state without the filament. \label{fig:freeen}}
\end{figure}

The free energy as a function of filament radius for various source voltages is illustrated in Fig. \ref{fig:freeen}(a).  The curves indicate that the filament can exist in a long-lived metastable state at $x=x_f$ [i.e. with the right minimum shallower than the left one at $F(x=0)$].  It becomes stable at relatively high voltage, $V>(h^2/CR_L)\sqrt{3W\rho /\pi r_0^3\kappa}$.  On the other hand, finite radius filaments become unstable at source voltages below,
\begin{equation}
V_0=18\sqrt{W\kappa\rho /\pi r_0^3}.\label{eq:Vh}\end{equation}
$V_0$ is defined by the conditions $\partial F/\partial r=\partial ^2F/\partial r^2=0$ and is presented by the curve labeled $0.3$ V in Fig. \ref{fig:freeen}(a).

At $V=V_0$, the steady state CF radius takes on its minimum value,
$$r_{min}= r_0\sqrt{2/H}=\sqrt{2\rho h/\pi R_L}.$$
The related filament resistance is a maximum, $R_{max}=\rho h/\pi r_{min}^2 =R_L/2$.  The characteristic holding current, $I_h$, (below which the CF disappears) can be defined as the current attained at the minimum source voltage $V_0$ [see the circuit in Fig. \ref{fig:model}(a)],
\begin{equation}\label{eq:Ih}
I_h=V_0/(R_{max}+R_L)=2V_0/3R_L.
\end{equation}
Along the same lines, the holding voltage is given by $V_h=I_hR_{max}$, which leads to, $V_h=V_0/3$ ($V_h$ is the voltage across the bulk of the filament and should not be confused with the source voltage $V_0$).

The metastable filament is predicted to be extremely long-lived at source voltages just slightly above $V_0$.  Indeed, as seen from Fig. \ref{fig:freeen}(a), the activation barrier separating the metastable minimum can be as high as $W_B\sim 10^8$ eV.  More quantitatively, close to $V_0$ the shape of the free energy is described by the expansion,
$$\delta F=\frac{1}{2!}\frac{\partial ^2 F}{\partial x\partial V}|_{V_0,x_{min}}\delta x\delta V+\frac{1}{3!}\frac{\partial ^3 F}{\partial x^3}|_{V_0,x_{min}}(\delta x)^3,$$
where $x_{min}=r_{min}/r_0$, $\delta x=x-x_{min}$ and $\delta V=V-V_0$ yielding,
\begin{equation}\label{eq:critreg}
r=r_{min}\left(1+\sqrt{\frac{\delta V}{4V_0}}\right), W_B=\frac{4hW}{r_0H}\left(\frac{\delta V}{V_0}\right)^{3/2}.
\end{equation}
The large barrier values are due to a large number of particles constituting the filament: $h/Hr_0 \sim hr_{min}^2/r_0^3\gg 1$.

As illustrated in Fig. \ref{fig:freeen}(b), the metastable nature of a steady state CF does not appear with the filament immediately upon creation [i.e. when the thermal contribution has not yet taken effect and the first term is excluded from Eqs. (\ref{eq:freeen}) and (\ref{eq:freeen1})].  At
\begin{equation}\label{eq:quasithr}
V>V_c=\frac{h}{r_0}\sqrt{\frac{3W\rho}{2\pi CR_Lr_0}},
\end{equation}
the stability of the newly created filament is maintained solely by the field.  While that interpretation implies a threshold voltage, it is quantitatively limited by the fact that the present model considers filament creation and disappearance in one step processes, neglecting the possibility of filament nucleation \cite{karpov2008}.

\begin{figure}[tb]
\includegraphics[width=0.35\textwidth]{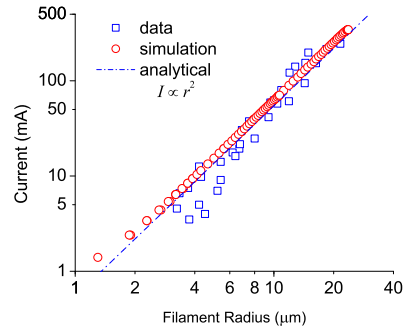}
\caption{(color online) Current as a function of CF radius indicating the typical $I\propto r^2$ dependence.  Excellent agreement is obtained between the analytical result of Eq. (\ref{eq:filrad}), our simulation, and data from Ref. \onlinecite{petersen1976}.\label{fig:tsplot2}}
\end{figure}

Beyond the critical region of voltage close to $V_0$, the filament radius becomes proportional to $\sqrt{I}$,
\begin{equation}\label{eq:filrad}
r=r_0\left(\frac{\rho h^2}{12\pi \kappa Wr_0}\right)^{1/4}\sqrt{I}\quad {\rm when}\quad I\gg I_h,
\end{equation}
consistent with the above mentioned experimental observations \cite{ovshinsky1968, petersen1976}, (see Fig. \ref{fig:tsplot2}).

\begin{figure}[tb]
\includegraphics[width=0.53\textwidth]{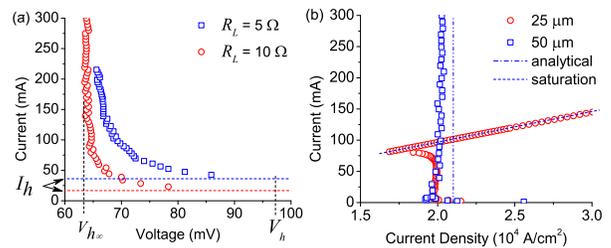}
\caption{(color online) (a) Simulated IV of a filament for the device described in Ref. \onlinecite{petersen1976} at two different load resistances $R_L$. The holding current $I_h$ and voltages $V_{h\infty}$ and $V_h$ are shown. (b) Filament current density for pores (active device regions) of diameters 25 $\mu$m and 50 $\mu$m. Saturation occurs when the CF fills the pore.  The results are in good agreement with the data \cite{petersen1976} (not shown here). \label{fig:tsplot3}}
\end{figure}

Eq. (\ref{eq:filrad}) predicts the often observed nearly vertical current voltage characteristic where the device voltage, $V_d=IR\propto I/r^2$, remains constant.  The value of this constant voltage at high current, $I\gg I_h$, is given by,
$$V_{h\infty}=V_0/3^{3/2}=V_h/\sqrt{3}.$$
The inequality $V_{h\infty}<V_h$ implies a `knee' in IV characteristic and a regime of negative differential resistance (NDR), as shown in Fig. \ref{fig:tsplot3}(a).

The reason for the apparently infinite dynamic conductivity $dI/dV\rightarrow \infty$ at $V\rightarrow V_{h\infty}$ is that, according to Eq. (\ref{eq:filrad}), the CF automatically adjusts its radius to maintain a constant current density,
\begin{equation}\label{eq:curden}
J=\frac{I}{\pi r^2}=\sqrt{\frac{12\kappa W}{\pi\rho r_0^3 h^2}}.
\end{equation}
It is a bulk material property and independent of current.  This  fact has been known empirically for more than four decades. When the CF grows to the device size, saturation is achieved and the current density increases linearly with current [see Fig. \ref{fig:tsplot3}(b)].

For the typical parameter values $\rho\sim 0.1$ $\Omega\cdot$cm, $\kappa\sim 10^{-3}$ cm$^2$s$^{-1}$, $h\sim 3000$ nm, $R_L\sim 100$ $\Omega$, $\varepsilon\sim 10$, and $A\sim 10^{10}$ nm$^2$  \cite{ovshinsky1968,petersen1976,adler1978,owen1973}, the numerical estimates for the above derived CF holding current $I_h\sim 1$ mA and current density $J\sim 10^4$ A/cm$^2$ are in excellent agreement with the data without any adjusting parameters. On the other hand, the predicted holding voltage $V_h\sim 0.3$ V is considerably lower than the measured $V_h\sim 1$ V. The latter discrepancy could be expected,  since our model here does not consider blocking electrodes \cite{petersen1976, adler1979} known to add a current-independent contribution to the voltage across device \cite{ovshinsky1968, adler1978}.

To complement our analytical work, finite element numerical simulations were performed using the COMSOL multiphysics package.  The electric field and temperature distributions were determined by simultaneously solving the coupled current continuity and heat equations, respectively. The results were then integrated to calculate the free energy of Eq. (\ref{eq:freeen}).  A search algorithm was used to determine the minimum free energy in the parameter space of applied voltage and CF radius.  The simulations relied on neither the flat-plate capacitor geometry nor the field and temperature approximations of Eq. (\ref{eq:deltaT}).

Numerous simulations were performed for a broad range of device sizes and geometries.  Figs. \ref{fig:tsplot2} and \ref{fig:tsplot3} provide samples of simulation results for the device structure described in Fig. 1 of Ref. \onlinecite{petersen1976}; overall, very good agreement was obtained without adjusting parameters.  The simulation results in Fig. \ref{fig:tsplot3}(a) clearly indicate the NDR knee in the bottom part of the IV curve, where the CF radius increases with $I$ much faster than $\sqrt{I}$.  That region of NDR corresponds to the critical region described in our analytical treatment [cf. Eq. (\ref{eq:critreg})].

In summary, we have developed a thermodynamic theory of steady state CF starting from the basic kinetic approach (Fokker-Planck equation). Analytical expressions have been derived and numerical model developed for the average CF characteristics. Our results correctly predict the filament properties observed in threshold switches and the corresponding features of their IV characteristics, particularly, negative differential resistance vanishing at high currents; the agreement with experimental data is remarkable.  Future work will extend this theory to include transient analysis and different applications (memory devices, shunting in thin-film photovoltaics, dielectric breakdown, etc.).

Useful discussions with Gianpaolo Spadini and Il'ya Karpov are greatly appreciated. We acknowledge the Intel grant supporting our research.

\end{document}